\documentclass[runningheads]{llncs}
\usepackage{graphicx}
\usepackage{caption}
\usepackage{multirow}
 \usepackage{amsmath}
 \usepackage{amsfonts}
\usepackage{url}
\begin{document}

\title{APS-USCT: Ultrasound Computed Tomography on Sparse Data via AI-Physic Synergy}
\titlerunning{APS-USCT}


\author{Yi Sheng\inst{1} \and
Hanchen Wang\inst{2} \and
Yipei Liu\inst{1} \and
Junhuan Yang\inst{1} \and
Weiwen Jiang\inst{1} \and
Youzuo Lin\inst{3} \and
Lei Yang\inst{1}
}
\authorrunning{Y. Sheng et al.}
%
\institute{George Mason University \\
\email{ysheng2@gmu.edu, lyang29@gmu.edu}\\
\and
 Los Alamos National Laboratory \and
 The University of North Carolina at Chapel Hill
}
\maketitle              
\begin{abstract}

Ultrasound computed tomography (USCT) is a promising technique that achieves superior medical imaging reconstruction resolution by fully leveraging waveform information,
outperforming conventional ultrasound methods. Despite its advantages, high-quality USCT reconstruction relies on extensive data acquisition by a large number of transducers, leading to increased costs, computational demands, extended patient scanning times, and manufacturing complexities. To mitigate these issues, we propose a new USCT method called APS-USCT, which facilitates imaging with sparse data, substantially reducing dependence on high-cost dense data acquisition. Our APS-USCT method consists of two primary components: APS-wave and APS-FWI. The APS-wave component, an encoder-decoder system, preprocesses the waveform data, converting sparse data into dense waveforms to augment sample density prior to reconstruction. The APS-FWI component, utilizing the InversionNet, directly reconstructs the speed of sound (SOS) from the ultrasound waveform data. We further improve the model's performance by incorporating Squeeze-and-Excitation (SE) Blocks and source encoding techniques. Testing our method on a breast cancer dataset yielded promising results. It demonstrated outstanding performance with an average Structural Similarity Index (SSIM) of 0.8431. Notably, over 82\% of samples achieved an SSIM above 0.8, with nearly 61\% exceeding 0.85, highlighting the significant potential of our approach in improving USCT image reconstruction by efficiently utilizing sparse data.

\end{abstract}
\section{Introduction}

Ultrasound Computed Tomography (USCT) is valued in the medical imaging landscape for its non-invasive nature and the absence of harmful radiation. This technique harnesses the potential of ultrasound data, which can be interpreted using either time-of-flight measurements (ray-based approaches)~\cite{Li-2009-Invivo} or full waveform data~\cite{Pratt-2007-Sound}. While ray-based USCT offers swift computational processing, it tends to compromise on image clarity. On the other hand, Full Waveform Inversion (FWI) significantly improves imaging quality by aligning synthetic waveform data with real recorded data, thus achieving a more detailed reconstruction of tissue sound-speed distributions than is possible with ray-based techniques. While FWI requires greater computational power, advancements in technology have made it more practical. The integration of cutting-edge machine learning and deep learning techniques with the power of high-performance computing has significantly improved USCT's effectiveness~\cite{li20213,lozenski2024learned}.

Although efficient, ML performance (i.e., the quality of SOS maps) relies on the highly dense waveform, which requires high-cost equipment with a large number of transducers (i.e., source and receiver).
In Figs.~\ref{fig:back} (a), (b), and (c), we elucidate the inverse relationship between data sparsity and the quality of the final SOS map reconstruction, demonstrating that SOS imaging quality diminishes as data sparsity increases.
Previous works \cite{jirik2012sound,guan2020limited} tried to address the data sparsity issue through model structure adjustment or algorithm optimization. However,
they have limited performance improvement since the root cause of low quality (i.e., the high sparsity of data) has not been addressed.

In this work, we propose to explore the feasibility of attaining high-quality reconstructed SOS maps by improving the given sparse data using an AI-physic synergy framework, namely 
APS-USCT.
The framework will first upscale sparse waveform by using an AI approach, called APS-wave, to generate the dense waveform.
To enable this, a training label dataset of dense waveforms is built by utilizing APS-physics.
Then, the generated dense waveform will be fed into the second AI component, namely APS-FWI, which uses the InversionNet as backbone architecture and 
integrates SE-Blocks \cite{hu2018squeeze} and source encoding \cite{wang2015waveform}. The SE-Block enhances detail capture in SOS map reconstruction, while the source coding boosts the model's learning efficiency. The main contributions of this paper are as follows:


\begin{figure*}[t]
  \centering  
  \includegraphics[width=1\textwidth]{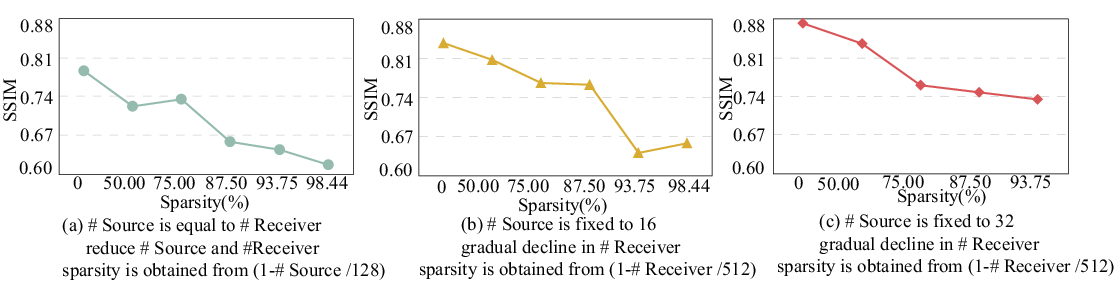}
  \caption{Relationship between sparsity and model performance}
  \label{fig:back}
\end{figure*}

\begin{itemize}
\item 
We propose an integrated framework, namely APS-USCT, to automatically reconstruct high-quality speed of sound (SOS) maps from sparse data.
\item In APS-USCT, the developed AI module (i.e., APS-wave) and physics module (i.e., APS-physic) work collaboratively to convert sparse measurements into dense waveforms to augment sample density prior to reconstruction, which can improve data
density while maintaining waveform integrity.
\item 
We evaluate APS-USCT on the breast reconstruction dataset, which outperforms the state-of-the-art techniques significantly. Compared with the state-of-the-art approach using dense input waveform, 
APS-USCT can achieve 
$2.5\times$ hardware cost reduction (i.e., less number of transducers) with merely 0.0007 SSIM degradation.

\end{itemize}

\begin{figure*}[t]
  \centering
  \includegraphics[width=1\textwidth]{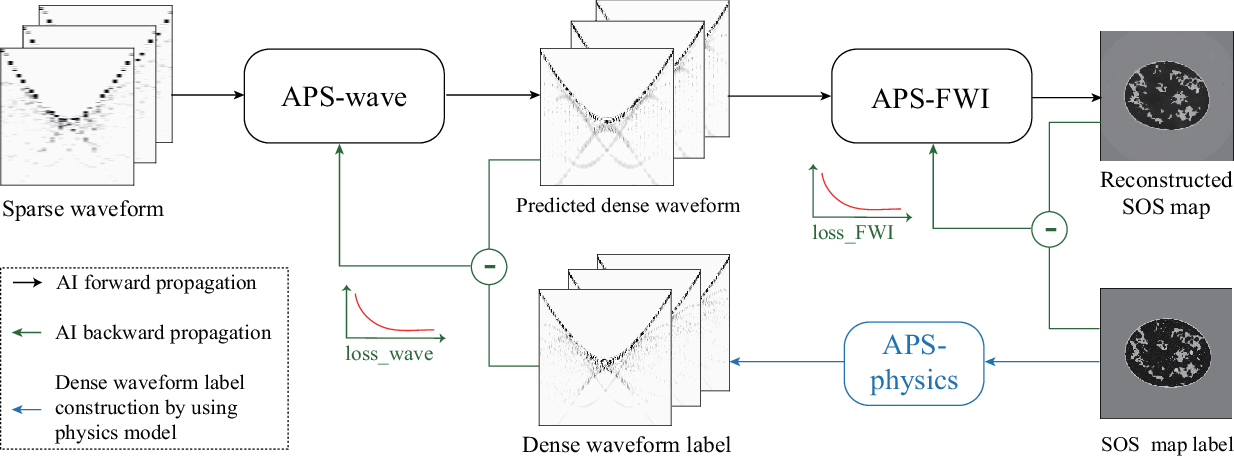}
  \caption{An overview of the proposed APS-USCT framework: (1) the black-color path shows the inference of reconstructing the speed of sound (SOS) maps from a given sparse measurement; (2) the green-color paths show the training procedure of two AI models associated with APS-wave and APS-FWI components; and (3) the blue-color path shows the waveform augmentation by using training APS-physics to generate the training dataset.}
  \label{fig:overviewframework}
\end{figure*}

\section{Method}

\noindent\textbf{\textit{A. Framework Overview:}} 
Fig. \ref{fig:overviewframework} shows the proposed framework, which is composed of three components: (1) APS-wave is an AI model that enhances sparse waveform data, increasing the density of samples (i.e., the sources or receivers), called ``dense waveform''; (2) APS-FWI is the other AI model for SOS map reconstruction from the dense waveform; (3) the underline APS-physics is the key enabler, which converts the SOS map label to dense waveform label.
As such, we have an AI-Physics Synergy USCT framework, denoted as APS-USCT.
In the following, we will introduce each component in detail.


\noindent\textbf{\textit{B. APS-wave:}}
The objective of APS-wave is to generate dense waveforms (i.e., higher sample density) from sparse ones obtained by fewer sources or receivers.
APS-wave contains two steps: (1) interleave the sparse waveform by inserting 0; as such, the interleaved sparse waveform will have the same dimension as the dense waveform; and (2) the interleaved waveform will be processed by a learnable encoder-decoder system which converts sparse measurements into
dense waveforms to augment sample density.

The encoder-decoder system contains the forward propagation and backward propagation, as shown in Fig. \ref{fig:overviewframework}.
For forward propagation, the interleaved waveform will go through a 15-layer encoder-decoder neural network to generate the dense waveform (please refer to the supplementary for the detailed structure).
For backward propagation, the key is to obtain the ground truth (i.e., dense waveform) for the given sparse waveform.
This is enabled by the APS-Physics (see later in this section).
Then, a Mean Squared Error (MSE) \cite{mse} loss function is applied between the predicted dense waveform and the ground truth, denoted as ``$loss\_{wave}$'' in Fig. \ref{fig:overviewframework}, which will be used to train the neural network.








\noindent\textbf{\textit{C. APS-FWI:}}
APS-FWI is to reconstruct the SOS map from the augmented waveform data, which contains two steps.
First, we integrate a learnable source encoding model, as described by \cite{wang2015waveform}, at the forefront of the InversionNet, and any other baseline methods for fair comparison. This model is tasked with encoding the predicted dense waveform through a random encoding vector. It aims to approximate the sound speed distribution via stochastic optimization with gradient descent. This process not only leverages the imaging operator's linearity for computational efficiency but also enables a significant reduction in computational demands.
Second, the encoded results will be fed to InversionNet \cite{wu2019inversionnet}, which aims to 
solve the minimization problem from a given waveform data.
\begin{equation}
\small
\min _{\boldsymbol{\xi} \in \mathbb{R}^{p}} \frac{1}{2 N} \sum_{n=1}^{N}\left\|\Phi_{\boldsymbol{\xi}}(\boldsymbol{D}^{n})-\boldsymbol{C}^{n}\right\|^{2}.
\end{equation}
where $\phi _{\xi}(D^{n})$ is the predicted SOS map on the $n^{th}$ sample by InversionNet $\phi$ with weights ${\xi}$ on input waveform data ($D^{n}$), and $C^{n}$ is the label of the $n^{th}$ sample.
The input $D^{n}$ is a 3-D tensor $D\in R^{I\times K\times J}$ and 
$d_{ijk}\in D$  corresponds to the measurement data from the $i^{th}$ source, $j^{th}$ receiver, and
$k^{th}$ time step.
The output $\phi _{\xi}(D^{n})$ and label $C^{n}$ are 2-D tensor $R^{x \times y }$ corresponding to pixel values of SOS estimates over the field of view.



Kindly note that the InversionNet was originally designed for seismic reconstruction in the geophysics domain, where the SOS maps describe the subsurface structures.
Although the fundamental FWI problem is the same, applying InversionNet to the medical domain encounters new challenges in reconstructing the details in the structure (e.g., tissue).
We propose to bring attention to InversionNet.
Specifically, we optionally add the Squeeze-and-Excitation (SE) Blocks in each layer of the decoder in InversionNet.
The experimental results will show that the InversionNet with attention (i.e., SE blocks) can better reconstruct tissue, particularly for sparse data.

\noindent\textbf{\textit{D. APS-physics:}}
To support the training process in APS-wave, the APS-physics is applied to generate high-dimensional waveform data from the SOS labels in the training dataset by using the acoustic wave equation \cite{wave}:


\begin{equation}
\small
\frac{\partial^2 u}{\partial t^2} = c(x)^2 \frac{\partial^2 u}{\partial \mathbf{x}^2} + S(t) \cdot \delta(\mathbf{x} - \mathbf{x_s^i}).
\end{equation}
where $u(\mathbf{x},t)$ represents the pressure wave field, \(c\) is the speed of sound (i.e., SOS) in the medium, \(t\) is time, and $\mathbf{x}$ is the spatial coordinates in the two-dimensional space. The term $S$ represents the source function, for which we employ a Ricker wavelet with 1MHz peak frequency, $\mathbf{x_s^i}$ represents the $i^{th}\ (i=1,2,...,N)$ source location coordinates, and $\delta(\cdot)$ is a spatial Dirac delta function. The recorded data at the $j^{th}\ (j=1,2,...,M)$ receiver position is described as $\boldsymbol{D}_j(t) = u(\mathbf{x}, t) \cdot \delta(\mathbf{x} - \mathbf{x_r^j})$, where $\mathbf{x_r^j}$ is the $j^{th}$ receiver location in space. This equation describes how the pressure wave \(u\) propagates over time in a constant density and isotropic medium. 
It is also considered the governing equation of a physics-based FWI algorithm for USCT.
Kindly note that by adjusting the number of sources $N$ and receivers $M$, we can use the above equation to generate pressure wave \(u\) with different sparsities.

%

\section{Experiment}

\subsection{Dataset, Implementation, and Evaluation Protocol}

To evaluate APS-USCT, we adopt 2D cross-sectional slices of SOS maps extracted from anatomically realistic numerical breast phantoms (NBPs),
which are constructed in \cite{li20213} by adapting and extending validated tools from the Virtual Imaging Clinical Trial for Regulatory Evaluation (VICTRE) project \cite{badano2018evaluation}
for use in USCT virtual imaging studies. A few examples of 2D cross-sectional slices extracted from these NBPs are available from \cite{2D}
Each NBP corresponds to a speed of sound (SOS) map, where the values were assigned randomly within realistic ranges, varying spatially.
The training dataset contains 1,353 NBPs while there are 41 NBPs in the testing set.

The AI and APS-physics in APS-USCT are implemented by using PyTorch and Python, respectively.
Specifically, the encoder-decoder architecture is adopted for APS-wave, InversionNet \cite{wu2019inversionnet} with additional SE-blocks for APS-FWI, and Forward Modeling algorithm \cite{forwardmodeling} for APS-physics.
Please refer to the supplementary for detailed training hyperparameters.


We apply two metrics for quantitative analysis: (1) structural similarity index measure (SSIM) \cite{wang2004image} that reflects the structure of objects in the scene, and (2) peak signal-to-noise ratio (PSNR) \cite{hore2010image} that is the approximate estimation of human perception of reconstruction quality.

We compare the quality of APS-USCT over three state-of-the-art methodologies: (1) InverstionNet \cite{lozenski2024learned}, (2) USCT-Net \cite{liu2021deep}, and (3) SRSS-Net \cite{long2023deep}.
Both USCT-Net and SRSS-Net are based on U-Net, and they are designed for sparse data. 
Unlike APS-USCT which generates dense waveforms, USCT-Net ensembles multiple low-quality SOS maps from waveforms taken by different sources. SRSS-Net uses an additional neural network to process the low-quality SOS maps, which is similar to the concept of superresolution in computer vision.
As InverstionNet did not consider data sparsity, we applied data interpolation methods for it, including (1) bicubic interpolation \cite{han2013comparison} (denoted Bicubi+InversionNet) and (2) nearest neighbor \cite{keller1985fuzzy} (denoted Nearest+InversionNet).

\subsection{Main Results}




\noindent\textbf{\textit{A. Quantitative and Qualitative Comparison:}} 
Table \ref{tab:main} reports the results for the waveform captured by 32 sources and 32 receivers, where the data sparsity is 93.75\% compared with the data used in InversionNet \cite{lozenski2024learned}.
We also report the proportion of test samples achieving SSIM over 0.8, 0.85, and 0.9.

As shown in the table, InversionNet, the baseline for comparison, recorded an SSIM of 0.7340 and a PSNR of 22.231.
It has 17.07\% of samples reaching SSIM of 0.8, but none reaching 0.85 or 0.9. 
Bicubic+InversionNet led to a lower SSIM of 0.6836, but Nearest+InversionNet increased the SSIM to 0.7603 from 0.7340
USCT-Net slightly outperforms InversionNet on SSIM by 0.0056, while it has a lower PSNR than InversionNet.
SRSS-Net can further improve the SSIM to 0.8089. 
Notably, both Nearest+InversionNet and SRSS-Net have reconstructed results over 0.85 for SSIM, but none of the existing approaches obtain SOS maps for SSIM over 0.9. 
Our proposed APS-USCT significantly enhances both the model's overall effectiveness and individual performance: the SSIM achieves a noteworthy level of 0.8431 and PSNR of 25.3040.
More importantly, there are 82.93\% of the test data achieve an SSIM above 0.8; moreover, 14.63\% of the test samples attain an SSIM exceeding 0.9.
These results underline the substantial improvement and value of APS-USCT that introduces to the breast USTC imaging reconstruction problem on space data.

\begin{table}[t]
\caption{ Comparisons between existing methods and APS-USCT}
\centering
\resizebox{\linewidth}{!}{%
\footnotesize
\begin{tabular}{c|c|c|c|c|c}
\hline
Method                & SSIM                    & PSNR                      & \begin{tabular}[c]{@{}c@{}}SSIM \\ \textgreater 0.8 (\%)\end{tabular} & \begin{tabular}[c]{@{}c@{}}SSIM\\ \textgreater 0.85(\%)\end{tabular} & \begin{tabular}[c]{@{}c@{}}SSIM \\ \textgreater{}0.9(\%)\end{tabular} \\ \hline
InversionNet  \cite{lozenski2024learned}        & 0.7340 (0.0120)         & 22.2312 (0.1931)          & 17.07                                                                 & 0                                                                    & 0                                                                     \\
Bicubic+ InversionNet & 0.6836 (0.0037)         & 20.6849 (0.1840)          & 4.88                                                                  & 0                                                                    & 0                                                                     \\
Nearest+ InversionNet & 0.7603 (0.0028)         & 22.6115 (0.0631)          & 29.27                                                                 & 4.88                                                                 & 0                                                                     \\
USCT-Net \cite{liu2021deep}                 & 0.7396 (0.0029)         & 18.5437  (0.4799)         & 17.07                                                                 & 0                                                                    & 0                                                                     \\
SRSS-Net  \cite{long2023deep}            & 0.8089 (0.0084)         & 23.5784 (0.7068)          & 56.10                                                                 & 21.95                                                                & 0                                                                     \\ \hline
\textbf{APS-USCT}     & \textbf{0.8431(0.0003)} & \textbf{25.3040 (0.0861)} & \textbf{82.93}                                                        & \textbf{60.98}                                                       & \textbf{14.63}                                                        \\ \hline
\end{tabular}
}
\label{tab:main}%
\end{table}

\begin{figure*}[t]
  \centering
  \includegraphics[width=0.98\textwidth]{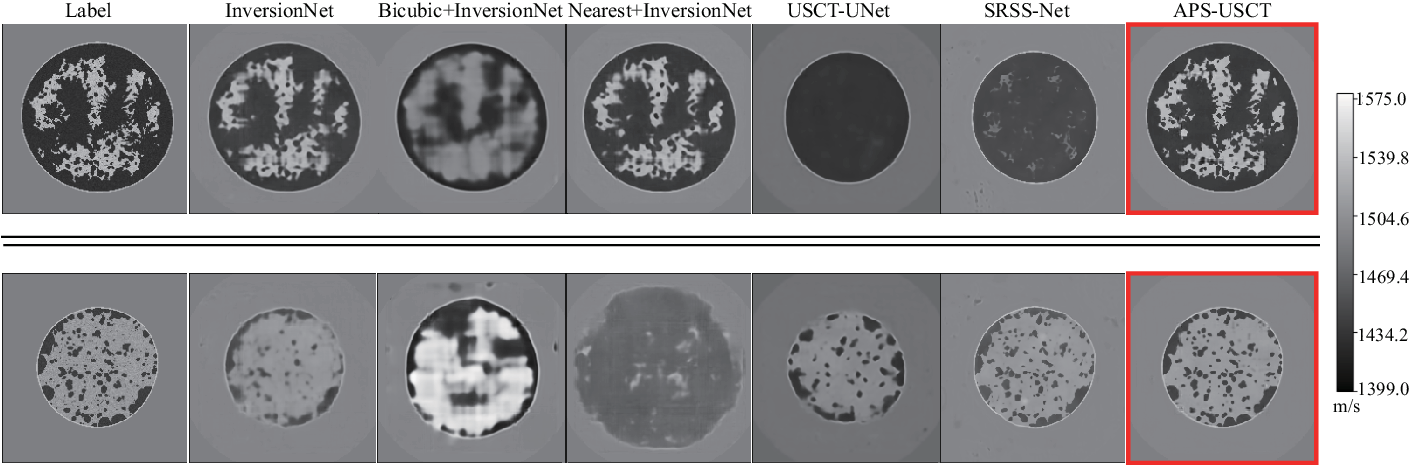}
  \caption{Visualization of the SOS map for the methods from the Table \ref{tab:main}}
  \label{fig:vis}
  
\end{figure*}

\noindent\textbf{\textit{B. Results Visualization:}} 
We visualize the SOS maps of 
obtained by different image reconstruction approaches in Fig. \ref{fig:vis}.
We selected two breast types (in terms of the percentage of fibroglandular tissue \cite{tissue,Birads}) for comparison: (1) fatty breast and (2) dense breast.
According to \cite{long2023deep}, dense breast tissue poses greater challenges in image reconstruction. 
We draw a crucial observation from the visualization results: for the existing models, a model with excellent overall average performance may still underperform on specific individual instances, which is even worse than a model has inferior overall performance.
In the upper section of Fig. \ref{fig:vis}, compared with the ground truth (i.e., label), 
Bicubic+InversionNet, USCT-Net, and SRSS-Net struggle to capture the SOS map's details.
In contrast, InversionNet, Nearest+InversionNet, and APS-USCT can reconstruct a clearer SOS map.
In the lower part of Fig. \ref{fig:vis}, we observe that only SRSS-Net and APS-USCT can capture the image information accurately.
This underscores the limitations of existing approaches in handling diverse breast tissues, with each method showing ineffectiveness on certain types.
Conversely, our method demonstrates superior robustness and effectiveness, consistently recovering detailed images across different tissue types.

\begin{figure*}[t]
  \centering
  \includegraphics[width=1\textwidth]{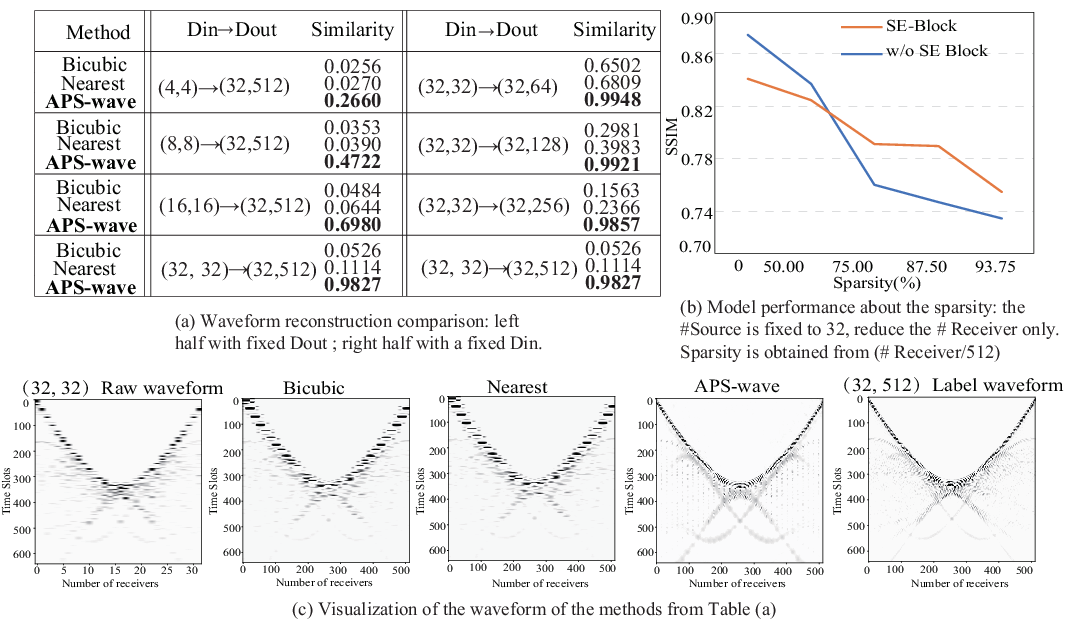}
  \caption{Results of the ablation studies for APS-wave and APS-FWI}    
  \label{fig:waveform_recon}
\end{figure*}

\subsection{Ablation Studies}

\noindent\textbf{\textit{A. Validity of APS-wave:}}
Fig. \ref{fig:waveform_recon}(a) reports the comparison of APS-wave over the interpolation methods.
The column $D_{in}\rightarrow D_{out}$ shows the change of (\# sources, \# receivers) pair for data acquisition.
The column ``similarity'' is the cosine similarity between the obtained data and ground truth by APS-physics.

We made several observations. 
First, traditional interpolation methods demonstrate limited efficiency. 
Both Bicubic and Nearest have similarity lower than 0.4, except $(32,32)\rightarrow(32,64)$, while APS-wave can achieve over 0.9 of similarity when $D_{in}$ is $(32,32)$.
Second, the higher similarity can be achieved by increasing $D_{in}$ or decreasing $D_{out}$.
For all methods, the best performance is achieved by $(32,32)\rightarrow(32,64)$, and the worst one is by $(4,4)\rightarrow(32,512)$.
Third, when $D_in$ is $(32,32)$, APS-wave demonstrates superior stability in increasing $D_{out}$, where it consistently achieves a similarity above 0.98.
But, others suffer a large similarity drop when increasing $D_{out}$ from $(32,64)$ to $(32,512)$.
All these results show the superiority of APS-wave in generating high-dense waveform. This is one major contribution that APS-USTC can outperform state-of-the-arts in Table \ref{tab:main}.

In Fig. \ref{fig:waveform_recon}(c), we visualize the waveform augmentation from $(32,32)$ to $(32,512)$.
The $(32,32)$ raw waveform exhibits clear breakpoints. 
We observed that these breakpoints remained largely unaddressed after applying Bicubic and Nearest techniques for waveform augmentation. 
This is because the interpolation methods simply expand the existing data points in the time domain and do not take the spatial relation of the source and receivers into consideration. 
In contrast, results from APS-wave exhibit a high degree of continuity,
which are close to the $(32,512)$ label waveform generated by APS-physics (most right-hand side one).

\begin{table}[t]
\centering
\caption{Resource-performance co-exploration of APS-USCT against InverstionNet \cite{lozenski2024learned} using dense data with higher hardware cost for training and inference}
\begin{tabular}{cccccc|cc}
\hline
\multicolumn{3}{c|}{InversionNet \cite{lozenski2024learned}} & \multicolumn{3}{c|}{APS-USCT} & \multicolumn{2}{c}{Comparison}                                                           \\
\multicolumn{1}{c}{$D_{raw}$}      & SSIM                 & \multicolumn{1}{c|}{\# Element} & $D$ by APS-wave &SSIM                 & \# Element             & SSIM Deg.            & HW Red.              \\ \hline
\multicolumn{1}{c}{(32,64)}        & 0.7468               & \multicolumn{1}{c|}{96}       & (32,32)$\rightarrow$(32,64) &0.7455               & 64                   & 0.0013               & 1.5$\times$                \\
\multicolumn{1}{c}{(32,128)}       &0.7602               & \multicolumn{1}{c|}{160}      &  (32,32)$\rightarrow$(32,128)&  0.7595               & 64                   & 0.0007               & 2.5$\times$                 \\
\multicolumn{1}{c}{(32,256)}       & 0.8369               & \multicolumn{1}{c|}{288}      &  (32,32)$\rightarrow$(32,256)& 0.8068               & 64                   & 0.0301               & 4.5$\times$                 \\
\multicolumn{1}{c}{(32,512)}       & 0.8734               & \multicolumn{1}{c|}{544}      &
(32,32)$\rightarrow$(32,512) &
0.8431               & 64                   & 0.0303               & 8.5$\times$                 \\ \hline
\multicolumn{1}{l}{}                & \multicolumn{1}{l}{} & \multicolumn{1}{l}{}          & \multicolumn{1}{l}{} & \multicolumn{1}{l}{} & \multicolumn{1}{l}{} & \multicolumn{1}{l}{}
\end{tabular}
\label{tab:tab2}%
\end{table}

\noindent\textbf{\textit{B. Effectiveness of SE-Block with APS-FWI:}}
Results in Fig. \ref{fig:waveform_recon} (b)
show that for input data with extremely high sparsity, say over 75\%, InversionNet with SE-Block can outperform the one without SE-block.
However, with high-density data, it becomes counterproductive.
Consequently, in APS-FWI, this module is optional—activated for sparse data. This approach provides more options for designers when they need to handle data with different sparsities.

\noindent\textbf{\textit{C. Resource-Performance Co-Exploration for APS-USCT:}}
Last, we conduct a set of experiments for resource-performance co-exploration on APS-USCT. 
As for comparison, we implement InversionNet \cite{lozenski2024learned} using dense data for both training and inference, indicating more hardware resources for data acquisition.
Here, the hardware cost is formulated as the number of elements used in the system, which is the sum of source numbers and receiver numbers.

The exploration results are reported in Table \ref{tab:tab2}.
We have several interesting observations from the table. 
First, with the same hardware cost, APS-USCT can continuously improve SSIM from 0.7455 to 0.8431 
when APS-wave generates a more dense intermediate waveform. 
This calls back to the results in Fig. \ref{fig:waveform_recon}(a), where APS-wave shows high stability.
On the other hand, for InversionNet, such improvements rely on increasing the number of hardware elements.
Second, the comparison between APS-USCT and InversionNet shows that APS-USCT can achieve $2.5\times$ hardware cost reduction with merely 0.0007 SSIM degradation.
When the hardware cost reduction increases to $8.5\times$, the SSIM degradation is still less than 0.031.
These results demonstrate that APS-USCT can significantly reduce hardware costs while maintaining high-quality reconstruction SOSs.

\section{Conclusion}

We developed the APS-USCT framework to improve image reconstruction from sparse ultrasound waveform, integrating AI with physical principles. APS-USCT features two key modules: APS-wave and APS-FWI. APS-wave improves sample density while maintaining waveform integrity. Following APS-wave, the APS-FWI module, enhanced by source coding and SE-Blocks, significantly improves reconstruction accuracy. This dual-module USCT method not only ensures an accurate tissue characterization with minimal hardware cost but also underscores APS-USCT's broad applicability in advanced ultrasound imaging technologies.



\subsubsection{\ackname} We express sincere gratitude to Dr. Mark A. Anastasio, Dr. Umberto Villa, and Dr.Fu Li for generously providing the dataset that was essential for this research.
We gratefully acknowledge the support of the National Institutes
of Health (NIH) (Award No. 1R01EB033387-01).

\subsubsection{\discintname}
The authors have no competing interests to declare that are relevant to the content of this article.



%




\clearpage
\bibliographystyle{splncs04}
\bibliography{paper.bib}






\clearpage
\newpage

\end{document}